# Growth of Zr/ZrO2 core-shell structures by Fast Thermal Oxidation

J.F. Ramos-Justicia*, J.L. Ballester-Andújar, A. Urbieta and P. Fernández
Department Materials Physics, Faculty of Physics, University Complutense
Plaza de Ciencias, 1, 28040 Madrid, Spain
*Corresponding author: juanfrra@ucm.es

## ABSTRACT

This research has been conducted to characterize and validate the resistive heating as a synthesis method for zirconium oxides ($ZrO_2$). A wire of Zr has been oxidized to form a core shell structure, in which the core is the metal wire, and the shell is an oxide layer around 10μm thick. The characterization This research has been conducted to characterize and validate the resistive heating as a synthesis method for zirconium oxides ($ZrO_2$). A wire of Zr has been oxidized to form a core shell structure, in which the core is the metal wire, and the shell is an oxide layer around 10μm thick. The characterization of the samples has been performed by means of Scanning Electron Microscopy (SEM). The chemical composition was analysed by X-ray spectroscopy (EDX). X-ray diffraction (XRD) and Raman spectroscopy have been used to assess crystallinity and crystal structure. Photoluminescence (PL) and cathodoluminescence (CL) measurements have allowed us to study the distribution of defects along the shell, and to confirm the degree of uniformity. The oxygen vacancies, either as isolated defects or forming complexes with impurities, play a determinant role in the luminescent processes. Colour centres, mainly electron centres as F, $F_A$ and $F_{AA}$, give rise to several visible emissions extending from blue to green, with main components around 2eV, 2.4-2.5eV and 2.7eV. The differences between PL and CL are also discussed.

.

**INTRODUCTION**

Transition Metal Oxides (TMO) are gaining relevance every day due to their wide range of applications, versatility, and advantageous physical properties (conductive, magnetic, luminescent, etc.), enabling the fabrication of multifunctional systems. However, for these applications to be really viable, a much deeper knowledge about critical factors such as carrier concentration, recombination rates, mobility, defect structure, and their influence on the band structure (and hence, on most relevant properties) is needed [1]. Applications in sectors as different as the Internet of Things (IoT), health or green energy demand high-performance materials. Sensors capable to detect the presence of toxic gases or to monitor changes in parameters critical for human health may be used to trigger a corrective or a warning protocol [2] ; more efficient catalysts might help to improve water cleaning or hydrogen production processes [3], which are crucial for attaining a cleaner environment and a safer, fair and more equal Society.

Some of these applications have a common bottleneck: the low efficiency of the redox reactions involved. The pre-requisite for a TMO to be efficient as a redox mediator is that the redox potential for the radicals or reactants involved lie within the band gap of the oxide (for instance in the case of photocatalysis, $E_o$ ($H_2O$/•OH) = 2.8V and $E_o$ ($O_2$/$O_2^{•-}$) = 0.28V vs. NHE [4]). In this context, the study of the defect structure and its influence on the band structure of the materials used become of the utmost importance. Moreover, a deep study of the effect of impurities and defects would pave the way to improve the tailoring capabilities. In this context, the role of composite materials is gaining ever more relevance since it opens the possibility to tune the band structure over a broader range, playing not only with the combination of materials but designing new architectures optimized for the different working conditions. The possible combinations of materials that can be made are numerous, but always the "active" part will be the semiconductor. Some common configurations are core-shell structures, and heterostructures or array of heterostructures ([5], [6]).

Zirconium oxide is a known TMO long ago, present in more than 40 minerals, mainly oxides and silicates. The most abundant zirconium minerals are zircon ($ZrSiO_4$) and baddeleyite ($ZrO_2$) and eudialyte ((Na, Ca)$_5$ (Zr, Fe, Mn)[O, OH, Cl][$Si_6O_{17}$]) the main source of this oxide is the mineral baddeleyite. Traditionally, the main uses of zirconium oxide have been those related its corrosion resistance, refractory properties and toughness, however more recently new applications have arisen and several review papers have been recently published [7]–[9]. Among the new applications found for $ZrO_2$, particularly in the micro-/ nanoscale, those for environmental remediation and biomedical applications are ever gaining relevance [9] . Considering the biomedical applications, antibacterial and antifungal activity play a central role. Basically, the same mechanisms are behind both antifungal and antibacterial applications. The first mechanism proposed is an electrostatic interaction between the negatively charged bacteria membranes and the $ZrO_2$ particles. The high surface area would favor this interaction and lead to the inhibition of the key metabolic functions, mediated by reactive oxygen species like oxygen and hydroxyl superradicals ($\cdot O_2^-, \cdot OH$) [10], [11]. A similar mechanism would operate in the case on antifungal activity inhibiting the cell division [12].

On the other hand, environmental remediation is probably the most pressing challenge faced by the present Society [13]. Removal of antibiotics, textile dyes and emerging pollutants as fluoride is becoming a first magnitude problem. The review paper published by Van Tran et al [9] contains a good revision of recent works on the use of zirconium oxide to eliminate different contaminants, with the added value that the $ZrO_2$ is synthesized by different green routes. Recently, Van Tran et al [9] have reviewed the different methods for green synthesis of zirconium oxide, using plant tissues and microorganisms as bacteria, algae or fungi.

Besides these health or environmental applications, zirconium oxide is gaining importance in several optoelectronic devices, and systems that require a high permittivity, as energy storage devices or Resistive Random Access Memories [14]. An important goal for optoelectronic applications is to optimize the doping processes. There is a general agreement on the relevant role of defects and dopants on the optoelectronic properties [9], [15]–[18]. Dopant may not only interact with intrinsic defects forming different complexes, but also they can modify dramatically the band structure of the material. $ZrO_2$ is a very wide bandgap material (>5eV) and consequently all the processes involving photoexcitation require ultraviolet illumination. The introduction of dopant as Ni causes a narrowing of the bandgap below 2.7e V [19] paving the way to the use of this material in micro-/ nanoelectronic devices.

In the present work we report on the fabrication of core-shell structures (Zr/$ZrO_2$) by thermal oxidation of the metal. Thermal oxidation occurs when an electrical current passes through a metal, Joule effect, while the metal is kept at ambient atmosphere. This procedure has been already proven to be effective in different oxides, being also suit for the growth of doped oxide shells [20], [21]. This method is simple and fast opening multiple possibilities to the fabrication of doped oxides and electronic composites.

**EXPERIMENTAL METHOD**

The experimental setup to grow the samples is shown in figure 1. A wire of the metal to be oxidized is suspended between two electrodes connected to a current source. The full system is kept at ambient atmosphere. As a consequence, of Joule heating, a temperature gradient is established along the wire, promoting the oxidation of the metal, giving rise to core-shell structures in which the core is the native metal, and the shell is an oxide layer about 10µm thick.

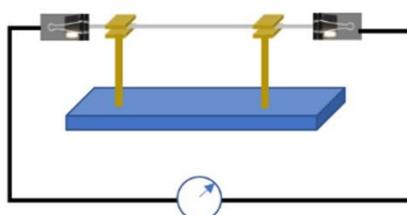

Figure 1. Experimental set-up for resistive heating growth

To study the influence of the growth parameters (current and time) different sets of samples have been prepared.

The starting material is a commercial as- drawn Zr wire (Goodfellows ZR00-WR-000130), with a nominal purity 99.2%. According to the provider´s specifications, the main impurity content corresponds to hafnium (2500ppm), other transition metals like chromium or iron are also present but at much lower concentrations (200ppm). The wire length (distance between electrodes is 10cm and the diameter is 0.25mm.

According to the temperature profile established in the wire, three different zones are defined for characterization purposes. Figure 2 shows a thermal image of the wire (a) and the zones defined (b). Thermal images are recorded with a FLIR E8-XT thermographic camera. The maximum temperatures reached are between 430 and 500ºC. Temperatures are measured with an optical pyrometer Infratherm IGA 12-S. There are some models which develop an analytical model for stationary distribution of temperature. A detailed explanation of the derivation of the differential equation that governs temperature profile because of Joule heating can be found in [22], whose notation is kept below:

$$\kappa A_{cr} T''(x) - (h\pi d - J^2 A_{cr} \rho_0 \xi) T(x) = J^2 A_{cr} \rho_0 (\xi T_0 - 1) - h\pi d T_0$$

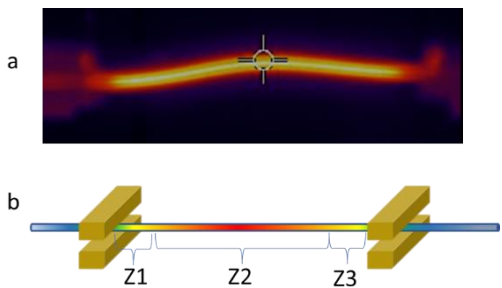

Figure 2.- a) Thermographic image of a wire. The cross indicates the point at which the maximum temperature is measured. b) Sketch of the three zones established for the characterization of the core-shell structures grown.

The general behaviour of the temperature profile can be written in terms of real exponentials (hyperbolic functions) or complex (sines and cosines), independent of initial conditions. Furthermore, this behaviour depends solely on the sign of the zeroth derivative temperature term. The sign of this term can be positive or negative at will by modifying physical parameters such as intensity, diameter, and resistivity (44 µΩcm (20°C for zirconium).. If this term vanished, a parabolic profile would be observed. In our case, the model does not adjust to experimental data, even including correcting terms as radiative heat loss to the prior equation. We believe that there might be size factors related to the wire diameter (0.25 mm) or physical processes that should not be disregarded in a first approach, what would make a difference on the accuracy of the model. A complementary, but extremely complex, model should account for the time evolution of the oxide shell. Nevertheless, this is out of our scope, since it would require a detailed knowledge of the growth kinetics, the defect structure at each growth step and their influence on parameters as the resistivity and thermal properties of zirconium oxide.

In this work we have prepared two different sets of samples. The first set consists of five wires submitted to currents from 2.2 to 2.8A for 10 seconds.

For the second set of samples, the maximum intensity has been 2.2A in all cases, but the current-time profiles and the times during which the wires are held at the maximum current have been varied. Figure 3 sketches the current –time profile used.

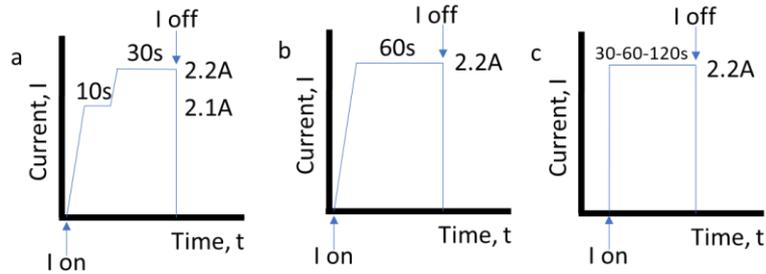

Figure 3.- Current time profiles used in the different experiments.

The characterization has been carried out by means of SEM based techniques and optical spectroscopy. Emissive mode images have been recorded in a FEI Inspect SEM. X- Ray microanalysis (EDX) has been performed in a Hitachi TM 3000 SEM equipped with a Brucker AXS Quantax system. μ-photoluminescence (μ-PL) and Raman spectroscopy measurements have been carried out in a confocal microscope Horiba Jobin-Ybon LabRAM HR800, with an excitation He-Cd laser source operated at 325 nm. Crystal structure has been studied by X-Ray Diffraction in a PANalytical Empyrean with the Cu-K$_\alpha$ line.

## RESULTS AND DISCUSSION

X-Ray diffraction patterns have been recorded under grazing angle, to avoid the signal of the Zr metal core. However, due to the morphology and low thicknesses of most of the oxide shells formed, the patterns are very noisy. For that reason, in figure 4

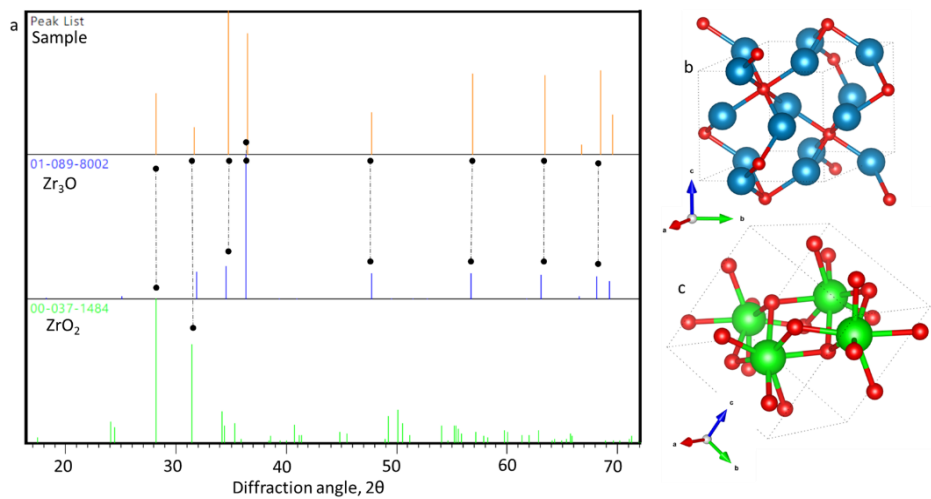

Figure 4.- a) XRD diffraction peaks obtained (upper) compared to those listed in the JCPDS cards for the corresponding phases: $Zr_3O$ (middle) and $ZrO_2$ (lower); b) Crystal structure of the $Zr_3O$ phase (O: red; Zr: blue); c) Crystal structure of the $ZrO_2$ phase (O: red; Zr: green)

we show the comparison between the main peaks of the pattern and the peak lists of the corresponding phases according to JCDPS cards.

Figure 4a (upper part) shows the peak list obtained from a typical X-ray diffraction pattern from our samples. The peaks fit with the presence of two zirconium oxides phases. The peaks shown in the middle graph correspond to a non-stoichiometric $Zr_3O$ hexagonal phase (JCDPS 01-089-8002) with lattice parameters a=5.6172Å, b=5.6172Å and c=5.1850Å (space group P63 2 2) (figure 4b). The $ZrO_2$ monoclinic phase is identified with the JCDPS 00-037-1484 card, with lattice parameters a=5.1473Å, b=5.2088Å and c=5.3166Å (space group P21/c) (figure 4c). No additional phases related to the presence of impurities are detected. The presence of non-stoichiometric phases could be expected since the oxidation process is extremely fast and might not be completed uniformly. On the other hand, this would give rise to areas of the sample with low crystal quality that would be responsible for the noisy patterns.

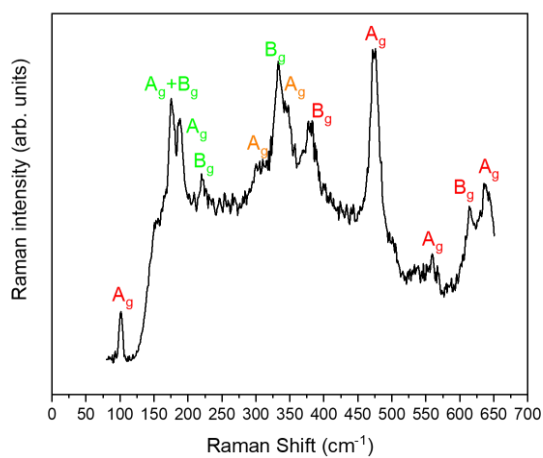

Figure 5. Typical Raman spectrum obtained from the samples. Coloured labels: red, O-O bonds; green, Zr-Zr bonds; orange, Zr-O bonds

To further assess the phase content, we have performed μ-Raman experiments. Then we can record Raman spectra from local regions, in particular from the central parts of the wires in which the shells are better formed, and the oxidation is more complete. The Raman spectra from all the samples studied are quite similar. The peaks observed correspond to the different $A_g$ (102, 188, 308, 346, 475, 561 and 637 cm$^{-1}$)and $B_g$ modes (175, 221, 334, 381 and 614 cm$^{-1}$) of monoclinic phase of the $ZrO_2$. The peaks are labelled according to the data found in the works of Quintard et al. [23] Ishigame et al. [24] and Kumari et al. [25]. All the peaks reported previously for the monoclinic phase are observed, and most of them are well defined and not much broadened, so the oxide shell is meant to have a good crystal quality. Red labels correspond to O-O bonds; the green ones, to Zr-Zr bonds; and the orange ones, to Zr-O bonds. The peak 175 cm$^{-1}$ is assigned to a combination of both $A_g$ and $B_g$ modes. Some modes (for instance that at 636cm$^{-1}$) could be also ascribed to the tetragonal phase, however since only monoclinic phase has been revealed through the X-Ray diffraction experiments, we discard this possibility and assign all the modes to the monoclinic structure.

The first set of samples described in the experimental method section has been used to check the influence of the temperature gradient established along the wire. From the point of view of the topography, no big differences are observed. Figure 6 shows typical images obtained from these samples. At low magnification (6a and b) a sort of columnar growth is observed, this morphology reflects the drawing tracks from the metal wire. This pattern is similar in all the samples although in those grown at higher intensities (b) the oxide layer is thicker and hence the grain structure is better defined. Part c of this figure shows a detail of the oxide layer, where the grains formed are clearly visible. The charge effects observed in this image are also a hint of the thickness of the oxide layer.

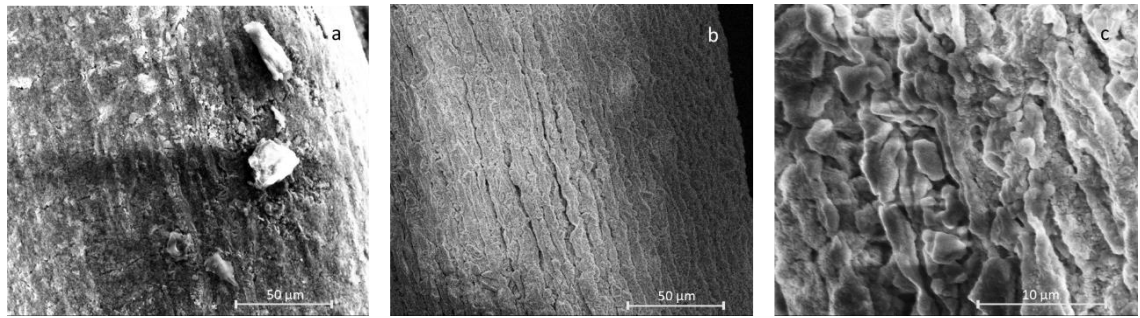

Figure 6.- SEM images obtained from simples grown at a current of: a)2.2 A; b) 2.8 A; c) detail of the sample shown in part b.

In the second set of samples, we have focused on the growth of microstructures and their evolution with treatment time. A current of 2.2A has been used in all cases; the difference resides on the time profile. In the samples grown with shortest times, (time profiles shown in figures 3a, 3b, and 3c with 30 seconds), only the first stages of microstructure growth are observed. As shown in figure 7, small needles grow between the grooves associated to the drawing lines.

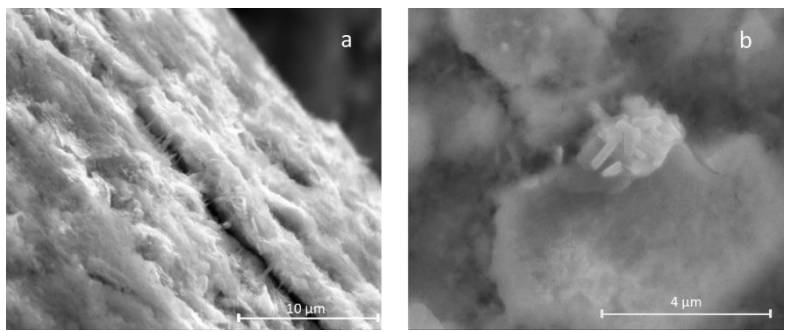

Figure 7.- SEM images of the nanostructures obtained at a current of 2.2A during 30s

The X-ray microanalysis shows a homogeneous distribution of Zr and O across the shell. Figure 8a shows a cross section of the wire. Figures 8b and c show the maps and line profiles corresponding to O and Zr respectively. The maps show clearly that oxygen is only detected from the outer part, that corresponds to the oxidized layer. The spectrum shown in figure 9 has been recorded from the shell (not in cross-section) and must be carefully considered. The spectra have been recorded at an accelerating voltage of 10KeV to minimize the contribution of the zirconium core to the signal. However, the $K_\alpha$ line from Zr (15.74keV) is not visible in these conditions and the characteristic line used for Zr is $L_\alpha$, with an energy of 2.04keV, but an intensity 25 times lower [26]. This estimation is based on the ratio of average intensity fluorescence coefficients of $K_\alpha$ and $L_\alpha$ shells [26]. That is the reason why apparently the most intense signals come from Si and S, whose $K_\alpha$ lines are placed at 1.739 and 2.307 keV respectively. According to the provider, the main impurities are Hf, O and C. Nevertheless, as shown in figure 9, Si and S are also observed, possibly added to Zr in the purification process from mineral ore. The carbon peak is due to graphite tape, used to stick the sample to the holder. The inset of figure 9 shows an estimation of the relative intensities of both Zr lines.

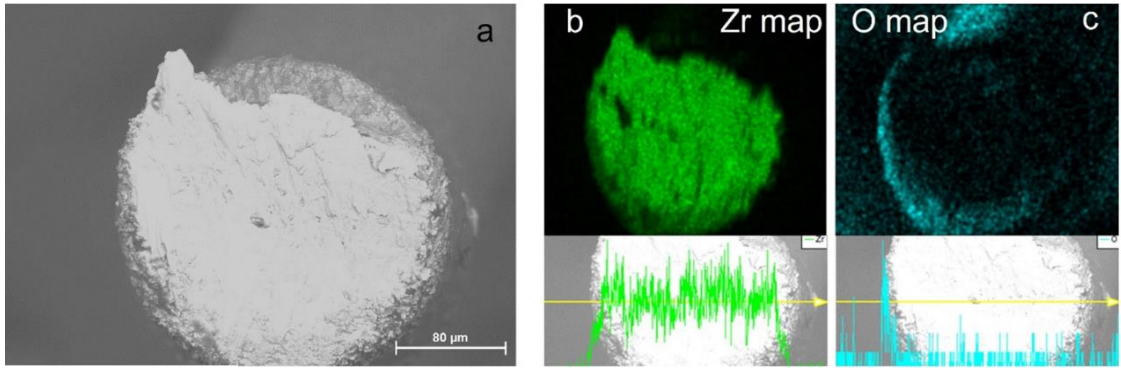

Figure 8: a) Cross section of the wire; b) Zr map upper and profile (lower) showing the uniformity of Zr distribution; c) O map (upper) and profile (lower), the oxygen signal is significant only at the outer surface of the wire (the lack of O signal at the right border is a shadow effect due to the topography of the sample and its orientation respect to the detector)

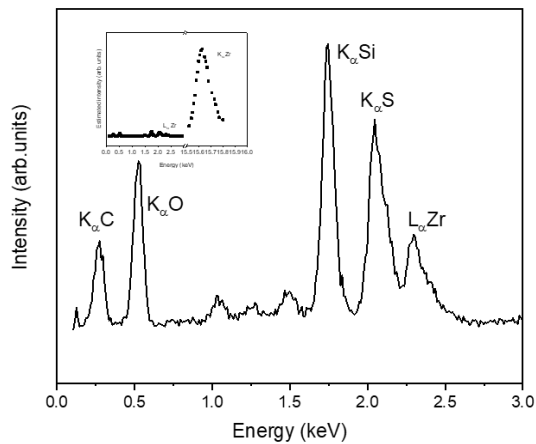

Figure 9.- EDX spectrum recorded with an accelerating voltage of 10keV. The inset shows an estimation of the intensity of the Zr $K_\alpha$ line, relative the peaks shown in the main graph.

Luminescence properties have been studied by means of photoluminescence (PL) and cathodoluminescence (CL). In both cases the spectra consist of a broad visible band extending from blue to green, although the relative intensity of the different components depends strongly on the excitation source. The origin of these emissions is not clear [27] but there is a general consensus about the role of defects and residual impurities on them [17], [28], [29]. The majority of defects found in oxides are colour centers. The most common are F- or $F_2$-type centers, oxygen vacancies or divacancies respectively, with trapped electrons. However T-centers constituted by two oxygen vacancies and a Zr cation ($V_O$-Zr-$V_O$) or hole centers as oxygen with a trapped hole ($O^-$), o V- type defects are also present [30]. The mechanisms behind the luminescence in $ZrO_2$ are not yet well understood, however, most of the previous works rely on the relation of the visible emission with oxygen vacancies and complexes $V_O$-M, where M is a cationic impurity that would occupy Zr sites [28], [29], [18] and references therein. In particular, we will refer to F, $F_A$ and $F_{AA}$ centers as the main responsibles for the visible emission bands.

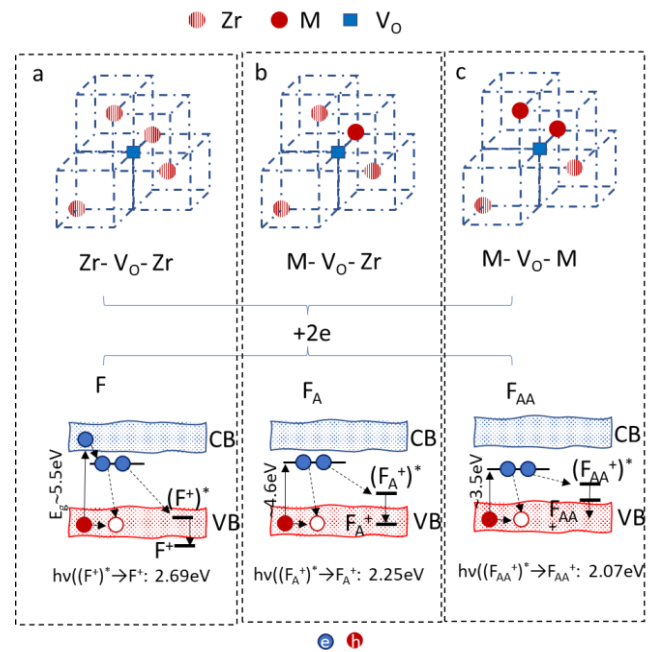

Figure 10.- Schematic representation of the different F centres expected in this system (refs 9-12) (upper) and the corresponding recombination paths (lower). a) Reactions 1, 3 and 4; b) Reactions 5 and 6; c) Reactions 7 and 8

Figure 10 shows a sketch of the three types of F centres present in the material. The F centre (part a) is formed when an oxygen vacancy adjacent to two Zr ions, traps two electrons [Zr- $V_O$- Zr]. The centres shown in parts b and c are extrinsic since they involve the cationic impurity. In the first case, (10b) Zr and the impurity cations are the nearest neighbours of the oxygen vacancy [M- $V_O$- Zr]. Finally, part c shows the defect formed when the two nearest neighbours are cationic impurities [M- $V_O$- M]. The most frequently reported impurities are $Ti^{+4}$ (or $Ti^{+3}$), as residual impurity that cannot be eliminated, and $Y^{+3}$ since yttria is one of the compounds used to stabilize the zirconia. According to the specifications of the provider, the main impurity present in the Zr wire is hafnium, that has been also mentioned as a candidate for the cationic impurity [31]. The defect reactions that would give rise to the visible emissions according to [32] are written below. The first step is the excitation of electrons through the band gap

$$ZrO_2 + h\nu\ (5.82eV) \rightarrow e^- + h^+ \quad (1)$$

Once the electron- hole pairs are created, several recombination paths could be activated. In our system, the electrons promoted to the conduction band could be easily trapped by the singly ionized oxygen vacancies ($F^+$ centres)

$$F^+ + e^- \rightarrow F \quad (2)$$

The trapping of a hole by the F centre would create an excited $F^+$ level

$$F + h^+ \rightarrow (F^+)^* \quad (3)$$

When the ground state is recovered

$$(F^+)^* \rightarrow F^+ + h\nu\ (2.69eV) \quad (4)$$

The defect reactions for the extrinsic F centres will be written as [18], [32], [33]

$$F_A^+ + h\nu\ (4.66eV) \rightarrow F_A + h^+ \quad (5)$$

$$F_A + h^+ \rightarrow (F_A^+)^* \rightarrow F_A^+ + h\nu\ (2.25eV) \quad (6)$$

$$F_{AA}^+ + h\nu\ (3.49eV) \rightarrow F_{AA} + h^+ \quad (7)$$

$$F_{AA} + h^+ \rightarrow (F_{AA}^+)^* \rightarrow F_{AA}^+ + h\nu\ (2.07eV) \quad (8)$$

The typical photoluminescence (PL) spectrum recorded under excitation with a 325nm He-Cd laser (figure 11a) consists of a broad band centered around 2.4eV. A similar band peaked around 2.4-2.5 eV has been previously reported [34]–[36] and attributed to oxygen vacancy related centers.

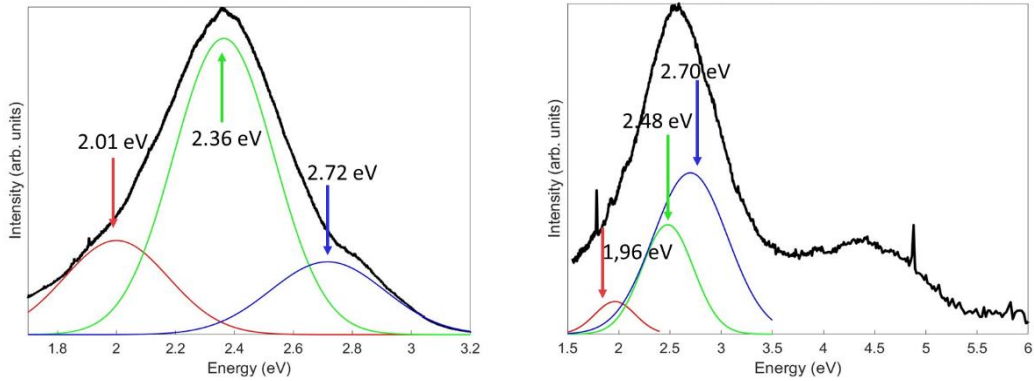

Figure 11.- Characteristic luminescence spectra: a) PL at an excitation wavelength of 325nm; b) CL at an accelerating voltage of 20keV

The deconvolution of this band in three components fits quite well with the transitions involving F centers mentioned above [33], [37]. It is worthy to note that the position of the electronic levels associated to defects are very sensitive to the surrounding and lattice relaxation state, that in our case could play an important role due to the low oxygen stoichiometry. The relative weight of the different components reflects the fact that the excitation source has an energy of 3.8eV (325nm) then electrons cannot be directly promoted to the conduction band, and F and $(F^+)^*$ must be filled directly with electrons from the valence band or different defect levels with lower energy. In our case, with an excitation of 3.8eV, the transitions involving $F_A^+$ and $(F_A^+)^*$ seem to be more favoured and hence the green emission predominates. Previous works [15]–[18], [31], [33], [37] report a broad blue band centered at 2.7eV, but either the excitation source are high energy photons (above the bandgap)[15], [16], [32], [33], [35], [36], [38]or they refer to cathodoluminescence experiments [18], [31], [34], [37]. The cathodoluminescence experiments performed in this work also point in this direction (Figure 11b). In this case the spectrum is centered closer to 2.6eV and the band is narrower than in the PL case. The deconvolution of the visible CL band shows three components at positions similar to those obtained in PL but with different relative intensities. In fact, the more intense component is now the blue one (2.7eV). In cathodoluminescence we are exciting the emission with high energy electrons (15-20keV) then, the electrons of the valence band may be easily promoted to the conduction band, and new and more efficient recombination paths could be favoured. The model proposed by Wang et al[37] attributes the 2.5eV emission to a transition involving the cationic impurity (Ti in the work of Wang). Electrons trapped by a shallow donor ($V_O^{\bullet}$), could be released either to the conduction band or to the F centers. If they are released to the CB, they could recombine with holes in the valence band, but also through the levels introduced by the complexes oxygen vacancy- cationic impurity, as depicted in the scheme from figure 12 adapted from [37].

Finally, the band observed in the CL spectrum centered at 4.5eV (figure 11b) could also be associated to this kind of complexes centers [37].

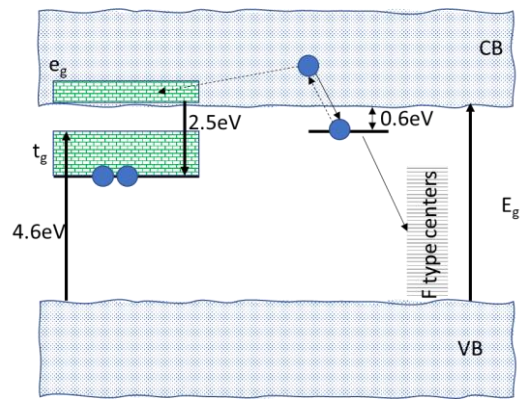

Figure 12.- Recombination paths in CL as proposed by Wang et al (16)

**CONCLUSIONS**

Zr/ZrO$_2$ core -shell structures have been obtained by rapid thermal oxidation of a Zr wire. The shell obtained seems to be quite uniform, except at the extremes where the temperature reached is lower. The zirconium oxide constituting the shell has a monoclinic structure although due to the lack of stoichiometry a Zr$_3$O phase has been also detected. This phase seems to have no major influence on the luminescent properties or crystal quality. Raman spectroscopy reveals the peaks characteristic of the monoclinic phase without a relevant broadening or shift with respect to the reported values. Luminescence spectra show a broad visible band that can be deconvoluted in three components at 2eV, 2.4-2.5eV and 2.7eV. All the emissions can be attributed to electron centres. In particular, different kind of F centres (oxygen vacancy related) let us to build a consistent scenario to explain the luminescent behaviour.

**ACKNOWLEDGMENTS**

**AUTHOR CONTRIBUTIONS**

Conceptualization, Ana Urbieta and Paloma Fernández; Formal analysis, Juan Francisco Ramos-Justicia, José Luis Ballester-Andújar, Ana Urbieta and Paloma Fernández; Funding acquisition, Paloma Fernández; Investigation, Juan Francisco Ramos-Justicia, José Luis Ballester-Andújar and Ana Urbieta; Methodology, Ana Urbieta and Paloma Fernández; Project administration, Paloma Fernández; Resources, Paloma Fernández; Supervision, Ana Urbieta and Paloma Fernández; Validation, Juan Francisco Ramos-Justicia, José Luis Ballester-Andújar, Ana Urbieta and Paloma Fernández; Visualization, Juan Francisco Ramos-Justicia, José Luis Ballester-Andújar, Ana Urbieta and Paloma Fernández; Writing – original draft, Juan Francisco Ramos-Justicia, José Luis Ballester-Andújar and Paloma Fernández; Writing – review & editing, Ana Urbieta and Paloma Fernández.